\documentclass[prb,aps,preprint]{revtex4}

\usepackage{graphicx}

\begin{document}

\title{Dynamics of a nanowire superlattice in an ac electric field}

\author{Aizhen Zhang}
\author{L. C. Lew Yan Voon}
\email{lok.lewyanvoon@wright.edu}
\affiliation{Department of Physics, Wright State University, 3640 Colonel Glenn Highway, Dayton, Ohio 45435, USA}

\author{M. Willatzen}
\affiliation{Mads Clausen Institute, University of Southern Denmark, Grundtvigs All\'{e} 150, DK-6400 S\o nderborg, Denmark}

\date{\today}

\begin{abstract}
With a one-band envelope function theory, we investigate the dynamics of a finite nanowire superlattice driven by an ac electric field by solving numerically the time-dependent Schr\"{o}dinger equation. We find that for an ac electric field resonant with two energy levels located in two different minibands, the coherent dynamics in nanowire superlattices is much more complex as compared to the standard two-level description. Depending on the energy levels involved in the transitions, the coherent oscillations exhibit different patterns.  
A signature of barrier-well inversion phenomenon in nanowire superlattices
is also obtained.
\end{abstract}


\maketitle

\section{Introduction}

Semiconductor quantum dots (QDs), also called ``artificial atoms,'' are
being considered for a variety of technological applications ranging from
semiconductor electronics to biological applications including optical
devices, quantum computing and biosensors. The study of coherent phenomena
is fundamental to a wide range of these applications. 
For example, a large number of experimental and theoretical investigations of Rabi oscillations (ROs), the
analog of atom-light coherent nonlinear interactions, have been
performed in the quantum-dot two-level systems in single QDs \cite
{Stie,Kama,Htoon,Zrenner,Borri,Beso,Mull,Wang,Li,Ulloa} and double QDs. \cite
{Haya,Petta}

If we deal with an array of QDs, the discrete levels broaden into energy bands. 
The question then arises whether coherent oscillations can be observed in such structures. 
The formation of minibands with discrete states for a system of fifteen coupled
QDs in a matrix has been demonstrated.~\cite{Kouwenhoven90}
Quantum transport in a model one-dimensional quantum dot array under a dc bias has also been studied 
theoretically.~\cite{Shangguan01} Another different type of QD array, so-called nanowire superlattices 
(NWSLs),~\cite{Wu,Gudiksen,Bjork,Solanki02,Wu04} has recently been fabricated using various approaches.
A nanowire superlattice consists of a series of interlaced nanodots of two
different materials. In NWSLs, the electronic transport along the wire axis
is made possible by the tunneling between adjacent dots, while the
uniqueness of each quantum dot and its zero-dimensional (0D) characteristics
are maintained by the energy difference of the conduction or valence bands
between the different materials. The band offset not only provides some amount
of quantum confinement, but also creates a periodic potential for carriers
moving along the wire axis. This new structure offers unique features and
suggests a diversity of possible applications, including nanolasers,
nanobarcodes, one-dimensional (1D) waveguides, resonant tunneling diodes,
and thermoelectrics.

The successful experimental developments of NWSLs have received increasing
theoretical attention. Recently considerable work has been devoted to NWSLs.
\cite{Lew,Galeriu,Lassen,Lin,Citrin,Madureira,Mizuno}
Lew Yan Voon and Willatzen \cite{Lew} have first studied the electron states and optical
properties in NWSLs. The thermoelectric properties of NWSLs have been
reported by Lin and Dresselhaus.\cite{Lin} Citrin \cite{Citrin} has
investigated the magnetic Bloch oscillations in nanowire superlattice rings.
Madureira{\sl \ et al.} \cite{Madureira} have proposed that dynamic
localization should be observable in NWSLs. A detailed study of acoustic
phonons in NWSLs has been given by Mizuno. \cite{Mizuno}

With the development of free-electron lasers that can be continuously tuned
in the terahertz (THz) range, the dynamics of charged particles in
semiconductor nanostructures has been a subject of intensive research. 
Motivated by the experimental progress and the importance to many applications
of NWSLs, 
in this work, we study the dynamics of a NWSL driven by an ac electric field as
a realistic system of a finite chain of coupled QDs. Diez {\sl et al. }\cite
{Diez} have investigated the dynamics of a finite semiconductor 
quantum well superlattice (SL) in an ac electric field and found that Rabi oscillations
between minibands are clearly identified under resonant conditions. However,
our system is different from SLs in a number of aspects. The coupling of
the superlattice longitudinal confinement to nanowire radial confinement in
NWSL leads to an additional position-dependent potential for the electron. 
Furthermore, for a nanowire superlattice, using a
position-dependent effective mass
leads to qualitatively new physics.~\cite{Lew}
Given these unique features, 
we expect that the dynamics of NWSLs driven by an ac electric field 
would reveal more interesting results.

The paper is organized as follows. In Sec. II we introduce the model and
present the theory. In Sec. III, we present the numerical results. A summary
is given in Section IV.

\section{Theory}

In this section we present the theoretical model and approach. The model is
essentially the same as the one used in Ref. \onlinecite{Lew}. We briefly describe
the model for completeness. The finite NWSL is modeled as an ideal cylinder surrounded by vacuum with
sharp modulations in the longitudinal (or $z$ ) direction (Fig.~1). The
electronic structure is obtained by solving the BenDaniel-Duke equation 
\begin{equation}
-\frac{\hbar ^2}2\nabla \cdot \left[ \frac 1{m({\bf r})}\nabla \psi ({\bf r}%
)\right] +V({\bf r)}\psi ({\bf r})=E\psi ({\bf r}), 
\end{equation}
where $V({\bf r)}$ is the potential experienced by the electron, $m({\bf r})$
is the effective mass in each layer, $\psi ({\bf r})$ is the wave function,
and $E$ is the energy. This model is appropriate for conduction states and
for large band-gap materials when nonparabolicity can be neglected. This
partial differential equation is separable in cylindrical circular
coordinates, leading to the following set of ordinary differential equations, 
\begin{equation}
\frac{d^2\Phi (\phi )}{d\phi ^2}+l^2\Phi (\phi )=0, 
\end{equation}
\begin{equation}
r\frac d{dr}(r\frac{dJ}{dr})+\left( q^2r^2-l^2\right) J(r)=0, 
\end{equation}
\begin{equation}
-\frac{\hbar ^2}2\frac d{dz}\left( \frac 1{m\left( z\right) }\frac{dZ\left(
z\right) }{dz}\right) +\left[ V\left( z\right) +\frac{\hbar ^2q^2}{2m\left(
z\right) }\right] Z\left( z\right) =EZ\left( z\right) , 
\end{equation}
where $\psi ({\bf r,}\phi ,z)$ $=$ $J(r)\Phi (\phi )Z\left( z\right) .$

The solution to the angular Eq. (2) is $\Phi (\phi )=e^{il\phi }$ with $l$
an integer. The solution to the radial Eq. (3) is a Bessel function of the
first kind, with the wavenumber $q$ determined by the boundary condition $%
J_l\left( qR\right) =0,$ where $R$ is the radius of the NWSL. Because the
zeros of the Bessel functions are themselves independent of the structure,
we only have to solve the longitudinal Eq. (4). Equation (4) describes an
electron moving in a one-dimensional effective potential $V^{eff}(z)=V\left(
z\right) +\frac{\hbar ^2q^2}{2m\left( z\right) }$. For typical
semiconductors, the well mass $m_W$ is smaller than the barrier mass $m_B.$
Thus the effective potential can be zero or negative if 
\begin{equation}
R^2\leq \frac{\hbar ^2\alpha ^2\left[ m_B(z)-m_W(z)\right] }{2V_B\left(
z\right) m_W(z)m_B(z)}=R_c^2,
\end{equation}
where $V_B\left( z\right) $ is the real barrier height and $\alpha $ is a
zero of the Bessel function. Hence, below a critical radius $R_c,$ the
barrier layer acts as the well layer and vice versa. The barrier-well
inversion induced by quantum confinement, a unique phenomenon in NWSLs, 
was predicted in Ref. \onlinecite{Lew}.

In order to study the dynamics of the NWSL driven by an ac electric field,
we must determine the band structure. At flatband, the band structure of Eq.
(4) is computed by using the finite-element method. 
The eigenstate $j$ of band $i$ with
eigenenergy $E_i^{(j)}$ is denoted as $Z_i^{(j)}(z).$ Under an ac electric
field, the envelope function for the electron wave packet satisfies the
following equation 
\begin{equation}
i\hbar \frac d{dt}\Psi (z,t)=\left\{ -\frac{\hbar ^2}2\frac d{dz}\left( 
\frac 1{m\left( z\right) }\frac d{dz}\right) +\left[ V\left( z\right) +\frac{%
\hbar ^2q^2}{2m\left( z\right) }\right] -eFz\sin (\omega _{ac}t)\right\}
\Psi (z,t),
\end{equation}
where $F$ and $\omega _{ac}$ are the strength and the frequency of the ac
field. Given the form of the initial wave function $\Psi (z,0),$ the
time-dependent Schr\"{o}dinger Eq. (6) is solved using the finite-difference
method and a fourth order Runge-Kutta integration. 
The mesh spacing is 0.1 nm and the time step is of the order $10^{-6}$ ps, 
which ensure the accuracy of the numerical calculation. For the sake of
simplicity, we have selected $\Psi (z,0)=Z_i^{(j)}\left( z\right) $ as the
initial wave packet. The probability of finding the electron in the state $%
Z_k^{(l)}\left( z\right) $ is given by 
\begin{equation}
P_{ik}^{(jl)}=\left| \int_{-\infty }^{+\infty }dzZ_k^{(l)}\left( z\right)
\Psi (z,t)\right| ^2.
\end{equation}

Note that, for the case of a two-level system with levels $a$ and $b$ driven
by a strong resonant field, the driving field couples levels $a$ and $b$ and
induces oscillations of the population in the two levels,
the so-called Rabi oscillations. The frequency $\Omega _{R}$ of the ROs is proportional to
the electric field amplitude and the dipole transition matrix element, which
is given by 
\begin{equation}
\Omega _R=eF_{Thz}\left\langle \psi _a\right| x\left| \psi _b\right\rangle
/\hbar =eF_{Thz}x_{ab}/\hbar ,
\end{equation}
where $F_{Thz}$ is the amplitude of the
terahertz electric field, $\left| \psi _a\right\rangle $ and $\left| \psi
_b\right\rangle $ are respectively the unperturbed eigenfunctions of levels $%
a$ and $b$, and $ex_{ab}$ is the dipole transition matrix element with $e$ the electric charge. If the
particle is initially in level $a$, the probability of finding the electron
in level $b$ oscillates between $0$ and $1$, i.e, complete resonance occurs.

\section{Results}

We consider a GaAs/GaAlAs NWSL structure with $10$ periods of $5$ nm GaAlAs
and $9$ periods of $10$ nm\ GaAs terminated by a 
cylindrical 100 nm GaAs layer on each side (Fig.~1). 
Such a structure can be well described by
Eq. (1). In the calculation we use parameters $\alpha =2.405$ (the first
root of zeroth order Bessel function), $V_z=230$ meV, $m_B(z)=0.0919m_0,$ and $%
m_W(z)$ $=0.067m_0$, where $m_0$ is the free-electron mass. The resulting
critical radius $R_c=2$ nm for such a structure. We first choose $R=5$ nm,
which is larger than the critical radius. We begin by considering the case
of resonant excitation with $F=50$ kV/cm. Figure 2(a) shows $P_{12}^{(55)}(t)
$ at the resonant frequency $\omega _{ac}=\left( E_2^{(5)}-E_1^{(5)}\right)
/\hbar =127$ THz, i.e., we are monitoring the transition between the
central state $j=5$ in the first miniband to the central state $l=5$ in the
second miniband. As can be seen, the oscillations reveal the occurrence of
coherent oscillations between minibands. 
This is qualitatively similar to the picture reported for one-dimensional superlattices.~\cite{Diez}
Figure 2(b) displays $P_{12}^{(22)}(t)$ at the
resonant frequency $\omega _{ac}=\left( E_2^{(2)}-E_1^{(2)}\right) /\hbar
=121$ THz, i.e, the transitions between the second state $j=2$ in the first
miniband to the second state $l=2$ in the second miniband. There are two
main differences in the oscillations in comparison to Fig. 2(a). Firstly,
the oscillation peak values are smaller. Secondly, the oscillations have a
decreasing amplitude tendency. This is due to the higher probability of the
electron tunnelling to the outer GaAs layers when the energy levels are close to
the boundary of the actual structure, 
because in the calculation we have considered an overall structure
much larger than the actual NWSL region. In Fig. 2(c)
we present $P_{12}^{(56)}(t)$ at the resonant frequency $\omega _{ac}=\left(
E_2^{(6)}-E_1^{(5)}\right) /\hbar =131$ THz, i.e., the probability of finding
the electron, initially in the central state $j=5$ in the first miniband, in
the state $l=6$ in the second miniband at time $t$. We find that coherent oscillations between
minibands can also be clearly identified except that the peak values 
are smaller than in Fig. 2(a). 
By performing the fast Fourier transform of $%
P_{12}^{(55)}(t)$, $P_{12}^{(22)}(t),$ and $P_{12}^{(56)}(t),$ we find that
the oscillation frequencies are $15.1$ THz$,$ $19.1$ THz, and $16.0$ THz,
respectively.

The different oscillation patterns and oscillation frequencies shown in Fig.~2 
can be understood as follows. The considered nanowire superlattice
structure presents Bloch minibands with nine states in each miniband. These
states are closely spaced with separations less than 1 meV. When a resonant
ac electric field is applied to two energy levels respectively located in
the first miniband and the second miniband, a number of energy levels are
involved in the transitions. As a result, the probability of finding the
electron in the energy level in the second miniband is much less than 1
(as shown in Fig. 2). To further check the validity of the two-level description, 
we calculate the oscillation 
frequencies corresponding to the three cases shown in Fig. 2
using the analytic Eq. (8) of a driven two-level system. The resulting
frequencies are $19$ THz, $4.9\times 10^{-2}$ THz, and 
$2.1\times 10^{-2}$ THz, respectively, which,
except for the first one, are very different from the values obtained by
performing the fast Fourier transform 
of $P_{12}^{(55)}(t)$, $P_{12}^{(22)}(t),$ and $%
P_{12}^{(56)}(t)$. Thus it is clear that the two-state description is not
generally valid for a NWSL. In the three cases shown in Fig. 2, different energy
levels and therefore different eigenfunctions are involved in the
transitions, leading to different oscillation patterns and oscillation
frequencies. Although the occurrence
of ROs has been identified in Ref. \onlinecite{Diez}, 
the important features of the 
sensitivity of ROs to the
energy levels involved in the transitions and the inapplicability of two-level 
theory were not pointed out.
Furthermore, because the electron can be driven up to the third miniband, we
find that the total probability of finding the electron in the second
miniband cannot reach 1 maximally.

We now turn to the case of nonresonant excitation. 
Figures 3(a)--3(c) show the transition probability $P_{12}^{(55)}(t),$ $P_{12}^{(22)}(t),$
and $P_{12}^{(56)}(t)$ with $\omega _{ac}=100$ THz and $F=50$ kV/cm
respectively. In contrast to the resonant situation, one can see that the
oscillation amplitudes are significantly decreased in all three cases when
the ac driving field is out of resonance.

Next we discuss the barrier-well inversion induced by quantum confinement
predicted in Ref. \onlinecite{Lew}. When $R< R_c,$ the barrier becomes the
well and vice versa. For the GaAs/GaAlAs NWSL structure discussed here, the
barrier width increases if the barrier-well inversion occurs, resulting in a
narrower band width and a larger band gap. The energy levels in the
minibands become more closely spaced and the wavefunctions change
dramatically. To compare with the case of $R> R_c$, in Fig. 4 we show
the calculated $P_{12}^{(55)}(t)$ with $R=1$ nm (less than critical radius).
Other parameters are the same as used in Fig. 2(a). As expected, the
oscillation pattern is totally different from Fig. 2(a) due to different
band structures. Both the amplitudes and the frequency are found to be
greatly decreased. Therefore the barrier-well inversion phenomenon is verified 
by the transport properties presented here.

\section{Summary}

In this paper, within a one-band envelope function theory
and by means of
numerically solving the time-dependent Schr\"{o}dinger equation, we have
presented an analysis of the dynamics of a finite NWSL driven by an ac
electric field. 
We have found that for the case of resonant ac electric field, the coherent dynamics 
is much more complex as compared to the standard two-level description. 
The oscillations are extremely sensitive to
the energy levels involved in the transitions. As a result, the oscillation
patterns are very different when the applied ac electric field is resonant
with two different energy levels respectively located in the first and second minibands.
Coherent oscillations between minibands can be identified when the probability
of the tunneling of the electron to the surrounding region is negligible. On
the other hand, the oscillations can also exhibit a decreasing amplitude
tendency due to the greater probability of the electron tunneling to the surrounding region. In
comparison to the resonant case, the oscillation amplitudes are
significantly decreased when the ac electric field is out of resonance. We
have also verified the existence of barrier-well inversion phenomena due to
the coupling of the superlattice longitudinal confinement to nanowire radial
confinement in NWSLs. This indicates that the radius of NWSL should be kept
in mind for investigating a variety of phenomena.

Nanowire superlattices do have some advantages
over other nanostructures: (i) scattering events are
highly suppressed because of the spatial confinement of carriers along all
three directions, leading to a long decoherence time, (ii) the wire
structure lacks the transverse excitations present in quantum-well superlattice
structures, eliminating this source of damping, and (iii) this novel
structure has no wetting layer (WL) compared with 
previously-studied single and double QDs.
These make NWSLs potentially more favorable for observations of coherent phenomena.
However, due to the structural complexity and the materials diversity in
these nanostructures, for practical applications and device optimization it
is essential to develop appropriate models to understand the behavior and to
predict properties of interest in these novel structures. Work along these
lines is currently in progress.

\begin{acknowledgments}
The work was supported by an NSF CAREER award (NSF Grant No. 0454849),
and by a Research Challenge grant from Wright State University and the Ohio
Board of Regents. 
\end{acknowledgments}


\newpage

\begin{figure}
\includegraphics[width=10cm]{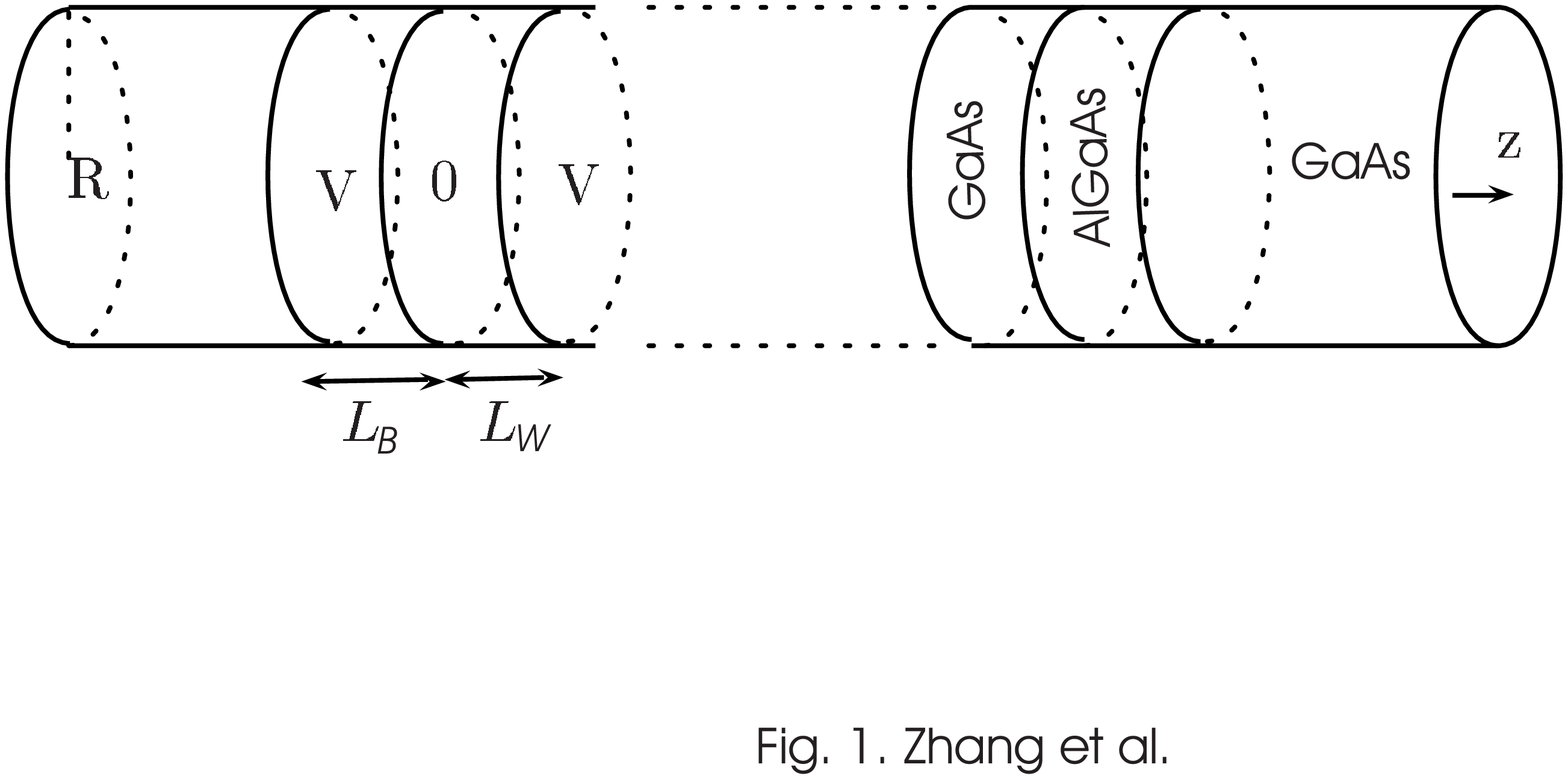}
\caption{\label{fig:fig1} 
Nanowire superlattice with radius $R$, well width $L_W$, barrier
width $L_B,$ and potential energy $V$.
The wire is capped on both sides by a thick cylindrical layer of GaAs.
}
\end{figure}

\begin{figure}
\includegraphics[width=10cm]{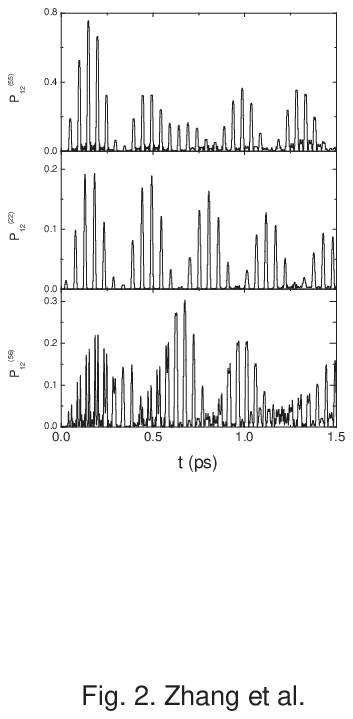}
\caption{\label{fig:fig2} 
(a) Temporal evolution of $P_{12}^{(55)}(t)$ with $\omega
_{ac}=\left( E_2^{(5)}-E_1^{(5)}\right) /\hbar =127$ THz. (b) Temporal
evolution of $P_{12}^{(22)}(t)$ with $\omega _{ac}=\left(
E_2^{(2)}-E_1^{(2)}\right) /\hbar =121$ THz. (c) Temporal evolution of $%
P_{12}^{(56)}(t)$ with $\omega _{ac}=\left( E_2^{(6)}-E_1^{(5)}\right)
/\hbar =131$ THz. In all cases, $F=50$ kV/cm and $R=5$ nm.
}
\end{figure}

\begin{figure}
\includegraphics[width=10cm]{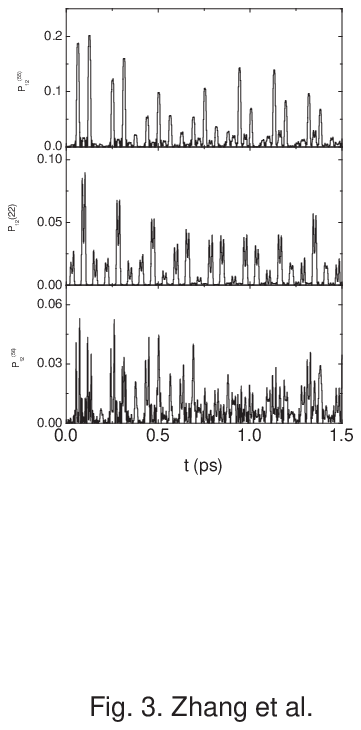}
\caption{\label{fig:fig3} 
Temporal evolution of (a) $P_{12}^{(55)}(t),$ (b) $P_{12}^{(22)}(t),$
and (c) $P_{12}^{(56)}(t)$ with $F=50$ kV/cm, $\omega _{ac}=100$ THz, and $%
R=5$ nm.
}
\end{figure}

\begin{figure}
\includegraphics[width=10cm]{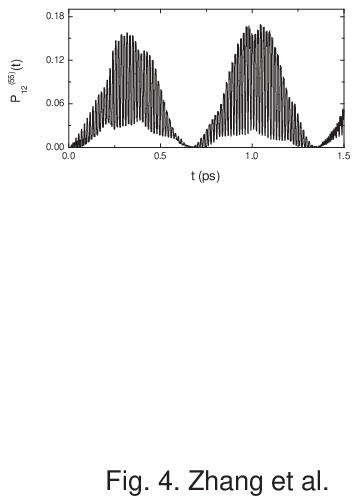}
\caption{\label{fig:fig4} 
Temporal evolution of $P_{12}^{(55)}(t)$ with $\omega _{ac}=\left(
E_2^{(5)}-E_1^{(5)}\right) /\hbar =127$ THz, $F=50$ kV/cm, and $R=1$ nm.
}
\end{figure}

\end{document}